\documentclass[a4paper,12pt]{article}
\usepackage{amsfonts, amsmath, graphics}
\usepackage[cp1250]{inputenc}
\usepackage[english]{babel}

\begin{document}
\begin{center}{\LARGE\bf Factor ordering in standard quantum cosmology}\end{center}

\begin{center}Roman \v{S}teigl, Franz Hinterleitner\\

{\em Department of Theoretical Physics and Astrophysics,\\
Faculty of Science, Masaryk University,\\
Kotl\'{a}\v{r}sk\'{a} 2, 611 37 Brno, Czech Republic.}\end{center}
 
\abstract{The Wheeler-DeWitt equation of Friedmann models with a massless scalar quantum field is formulated with arbitrary factor ordering of the Hamiltonian constraint operator. A scalar product of wave functions is constructed, giving rise to a probability interpretation and making comparison with the classical solution possible. In general the behaviour of the wave function of the model depends on a critical energy of the matter field, which, in turn, depends on the chosen factor ordering. By certain choices of ordering the critical energy can be pushed down to zero.}

\section{Introduction}
We consider homogenous, isotropic cosmological models with a scalar, massless field as source. Both the gravitational degree of freedom and the scalar field are quantized, thus leading to a simple quantum mechanical system. The notoriously ambiguous factor ordering in the Hamiltonian constraint operator in the Wheeler-DeWitt equation is left open by introducing a two-parameter family of factor orderings. In all the cases we arrive at a Bessel or a modified Bessel equation as wave equation for negative and positive curvature, respectively. A probability interpretation of the wave functions is supported by considerations about the possibility to derive an equation of continuity from the Wheeler-DeWitt equation, which is influenced by the factor ordering. Then for not too small values of the scale factor of the cosmological model the probability densities, derived in a natural way from the wave functions, approach their respective classical limits.   

An analysis of the factor ordering issue in closed Friedmann models can be found in \cite{KW}. That work is aimed at a critical review of the Hartle-Hawking no-boundary \cite{HH} and the Vilenkin tunneling wave functions \cite{V} of the universe. In the meantime progress has been made in Loop quantum cosmology based on a quantum theory of gravity \cite{LQC}, which allows to calculate and make physical sense of wave functions at and beyond the classical cosmologic singularities and where the factor ordering plays a key role, too. Whereas in \cite{KW} the potential of the scalar field is approximately constant and its kinetic energy does not play a role in the Wheeler-DeWitt equation, we assume a free, massless field, only coupled minimally to gravity, leading to separable, exactly solvable equations. By introducing two real  parameters (and in one case complex ones) we consider a larger degree of freedom of factor ordering than \cite{KW}, where one integer parameter is used.

In all the cases, i.e. for the closed, the open, and the spatially flat model, there is a critical energy of the matter field with the associated gravitational wave function being regular (or going to infinity) near the singularity for energies below it and oscillating infinitely many times for energies above it.  The existence of a critical energy of the matter field is also observed in the framework of Loop quantum cosmology \cite{MB}. Here, in the simpler formalism of standard quantum cosmology, where we have reduced the system to a mechanical one before quantizing, but introduced factor ordering parameters, we can control the factor ordering by varying the parametrs arbitrarily. By variation of the factor ordering it is possible to shift the critical energy, it is also possible to choose such orderings, where the critical energy goes down to zero. So, if one would prefer a unique behaviour of the wave functions at the singularity at all energies, e.g. to satisfy a DeWitt condition of zero probability at the singularity, this would restrict the possibilities of factor ordering.   

\section{The classical model}
We start with the action
\begin{equation}
S=\int  c\left(\frac{1}{\kappa}L_{G}+L_{M}\right)\,\mathrm{d}\Omega
\end{equation}
where $L_M$ is the lagrangian for a scalar field $\phi$,
\begin{equation}
L_{M}= - (\phi_{;\mu}\phi_{;\nu}g^{\mu\nu}+V(\phi)),
\label{action}
\end{equation}
and $L_G$ is the geometrical lagrangian. Greek indices are spacetime, $V(\phi)$ is the potential of the scalar field. Under the assumption of homogeneity and isotropy a standard simplification of variables leads to the Robertson-Walker metric (RW metric) and after integration of the action (\ref{action})  over a unit volume\footnote{The Ricci scalar $R$ is only a function of the time for the RW metric, so only $\sqrt{ -g}$ is a function of spatial variables. In the closed case we can integrate over the whole volume, in the open case we must integrate over some finite, spatially constant ``test volume", so we can study time dependence of the scale for this volume. After this step, the lagrangian is only a function of time and the equations of motion deduced from this lagrangian by variation are equivalent with the Einstein equations for the RW metric. With these assumptions we have finite dimensional mechanical lagrangian and hamiltonian theory for the RW models and we can use this in quantum theory. In this example the reduction to a mechanical system before quantization works, but not generally \cite{MC}.}, we get the hamiltonian by a straightforward Legendre transformation, 

\begin{equation}
H= N \left[-6\left(\frac{1}{144}\frac{p_{a}^{2}}{a} + \tilde{k}a\right)+\frac{1}{4}\frac{p_{\phi}^{2}} {a^{3}}+ a^{3}V(\phi)\right].
\label{HV}
\end{equation}

Here $N$ is the lapse function, $a$ the scale parameter, $c$ the speed of light, $\kappa$ Einstein's gravitational constant $\kappa=\frac{8\pi G}{c^2}$ and $\tilde{k}$ is the normalized sectional curvature, which can be $+1,-1$ and $0$ for the closed, open and flat case. Momenta are defined in the standard way

\begin{eqnarray}
p_{a} &:= & \frac{\partial L}{\partial \dot{a}} = - 12 \frac{a \dot{a}}{N}\\
p_{\phi} &:= & \frac{\partial L}{\partial \dot{\phi}} = 2 \frac{a^{3} \dot{\phi}}{N}.
\end{eqnarray}

The set of Einstein's equations following from the action (\ref{action}) is nonlinear, which is a good reason for using the hamiltonian formalism for the solution of this problem. In the massless case the equations are

\begin{eqnarray}
\dot{p}_a&=&-\frac{1}{24}\left(\frac{p_a}{a}\right)^2+6\tilde{k}+\frac{3}{4}\left(\frac{p_{\phi}}{a^2}\right)^2,\\
\dot{a}&=&-\frac{1}{12}\frac{p_a}{a},\\
\dot{p}_{\phi}&=&0,\\
\dot{\phi}&=&\frac{1}{2}\frac{p_{\phi}}{a^3}.
\label{dyn4}
\end{eqnarray}
Together with the hamiltonian constraint
\begin{equation}
H=0
\label{con}
\end{equation}
we have a system equivalent with the Einstein equations for the RW metric.
\section{Classical solutions}
\subsection{Solution for $\tilde{k} =1$}

The momentum of the massless scalar field is constant. Thus one can write condition (\ref{con}) in the form

\begin{equation}
p_a = \pm \,12\,a\, \sqrt{\left( \frac{a_m}{a}\right)^4 -1},
\label{con2}
\end{equation}
where $a_m$ is the maximal radius of the universe, given by $a_m^4 = \frac{p_{\phi}^2}{24}$ and obtained for $\dot a=0$. If we combine condition (\ref{con2}) with the dynamical equations (6) and (7), we get
\begin{eqnarray}
\dot{p_a}&=&12 \left(\left(\frac{a_m}{a}\right)^4+1\right),\\
\dot{a}^2&=&\left(\frac{a_m}{a}\right)^4-1.\label{apunkt}
\end{eqnarray}
The expansion speed $\dot a$ in the second equation is the first integral of the first equation (if we put $p_{a}= - 12 a \dot{a}$).
 The second equation leads to the integral 
\begin{equation}
\int_0^a \left(\frac{\tilde{a}}{\tilde{a}_m}\right)^2 \frac{\mathrm{d}\tilde{a}}{\sqrt{1-\left(\frac{\tilde{a}}{\tilde{a}_m}\right)^4}}=\pm \int_{t_0}^{t} \mathrm{dt},
\label{integral}
\end{equation}
where we suppose $0\leq \tilde{a} < \tilde{a}_m$. The solution can be written in form of a Gau\ss\ hypergeometric function \cite{Abramowitz}
\begin{equation}
\pm (t-t_0) = \frac{a_m}{3}\left(\frac{a}{a_m}\right)^3 {}_{2} F_1 \left(\frac{1}{2},\frac{3}{4};\frac{7}{4};\left(\frac{a}{a_m} \right)^4 \right)\equiv \tilde{F}_c(a).
\label{sol}
\end{equation}

For the inversion of this function, leading to $a(t)$, we have to consider the expanding and the contractiong branch seperately. Denote the time of the maximal radius as $t_m =\tilde{F}_c(a_m)$. The expanding branch is obtained by choosing the positive sign on the l. h. side of (\ref{sol}) and $t_0=0$, whereas the negative sign and $t_0 = 2t_m$ yields the contracting branch. After the unification of the two branches we get the typical time evolution of the closed cosmological model without cosmological constant shown in figure 1, i. e. the time evolution from the big bang to the collapse. For completeness, the time of the maximal radius is in SI units
\begin{equation}
ct_m = \frac{a_m}{3}\frac{\Gamma(\frac{7}{4}) \Gamma(\frac{1}{2})}{\Gamma(\frac{5}{4})}\approx 0,599 \; a_m .
\end{equation}
Our exact solution is an explicit confirmation a theorem in \cite{Burnet} for the case of the massless scalar field.

\begin{figure}[h]
\centering
\rotatebox{-90}{
\scalebox{0.50}{%
\includegraphics{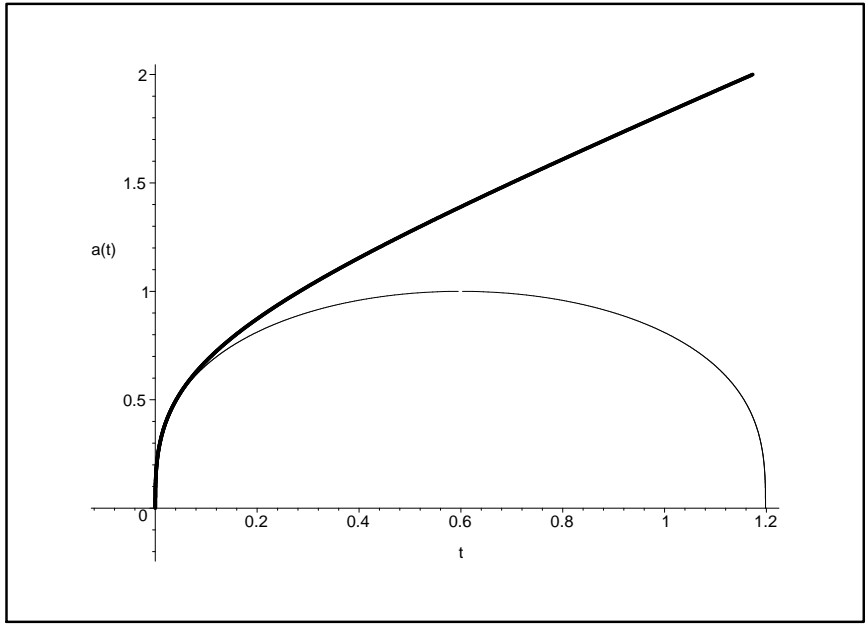}%
}}\caption{The time evolution of the RW models.}
\label{classpic}
\end{figure}
On figure \ref{classpic} the evolutions of the scale $a$ for the closed and the open (bold line) model, with $a_m=1$, $p=1$ (see definition below eq. (\ref{p})) and $c=1$ are displayed.
\par
Note that (\ref{sol}) is a solution of a Gau\ss\ equation which has second linearly independent solution: a constant.
The solution for the field is 

\begin{equation}
\phi=\phi_0 \pm \sqrt{\frac{3}{2}}\; \mathrm{arctanh}\sqrt{1-\left(\frac{a}{a_m}\right)^4},
\end{equation}
so we have fully described our system on the classical level.

\subsection{Solutions for $\tilde{k} =-1$ and $\tilde k=0$}
The open case is analogous. The only difference is a plus sign under the square root in (\ref{con2}) and in the integral (\ref{integral}) and of course there is no maximal radius of the universe.  The solution is 

\begin{equation}
\pm (t-t_0) = \frac{p}{3}\left(\frac{a}{p}\right)^3 {}_{2}F_1 \left(\frac{1}{2},\frac{3}{4};\frac{7}{4};-\left(\frac{a}{p} \right)^4 \right)\equiv \tilde{F}_o(a),
\label{sol2}
\end{equation}

\begin{equation}\label{p}
\phi=\phi_0 \pm \sqrt{\frac{3}{2}}\; \mathrm{arctanh}\left(1+\left(\frac{a}{p}\right)^4\right)^{-1/2},
\end{equation}
where $p^4 = \frac{p_{\phi}^2}{24}$ is the formal analog of $a_m$. If we choose $t_0=0$ for the positive sign in (\ref{sol2})  we get a solution for the scale parameter increasing in time. In this solution the expansion speed of the universe is converging in time to a constant, this follows from the equation 
\begin{equation}\label{minus1}
\dot{a}^2=\left(\frac{p}{a}\right)^4+1.
\end{equation}

Because $ a $ is increasing forever, it follows from this equation that $\dot a\approx 1$ for large $t$. 
This is typical for open cosmological models with vanishing cosmological constant. 

In the flat case $\tilde k=0$ we obtain the simple solution 
\begin{equation}
a(t)={\rm const.}\,^3\!\!\!\sqrt{t}.
\end{equation}

\section{Wheeler-DeWitt equation}
One of the possibilities how to choose a representation by selfadjoint operators on $\mathcal{L}^{2}(\mathbb{R}^+,\frac{da}{a};\mathbb{R},d\phi)$ is \cite{isham} 

\begin{equation}
\hat{a}\psi(a,\phi):=a\psi(a,\phi) \quad\quad \hat{\pi}\psi(a,\phi):= - \mathrm{i}\hbar a\frac{\partial\psi(a,\phi)}{\partial a} 
\label{nso}
\end{equation}
\begin{equation}
\hat{\phi}\psi(a,\phi):=\phi\psi(a,\phi)  \quad\quad \hat{p}_{\phi}\psi(a,\phi):= - \mathrm{i}\hbar\frac{\partial\psi(a,\phi)}{\partial \phi}. 
\end{equation}
For the motivation for this (noncanonical) choice of $\hat\pi$ and the above Hilbert space, in which it is selfadjoint, see \cite{isham} and quotations therein.

The Poisson brackets lead to the affine commutation relations

\begin{equation}
\{a,\pi \}=a \quad \rightarrow \quad [\hat{a},\hat{\pi}]=\mathrm{i}\hbar\hat{a},
\end{equation}

\begin{equation}
\{\phi,p_{\phi}\}=1 \quad \rightarrow \quad [\hat{\phi},\hat{p}_{\phi}]=\mathrm{i}\hbar,
\end{equation}
where classically $\pi=ap_a$.

With the usual substitution $p_a\rightarrow -i\hbar\partial/\partial a$ in the hamiltonian (\ref{HV}) we obtain the following two-parameter family of Wheeler-DeWitt equations\footnote{All quantum formulae are in SI units.}

\begin{equation}
\left(\frac{1}{24}\frac{\kappa \hbar ^2}{c}\frac{1}{a^{i}} \frac{\partial}{\partial a} \frac{1}{a^{j}} \frac{\partial}{\partial a}  \frac{1}{a^{k}} -  \frac{1}{4}\frac{\hbar ^2}{c}\frac{1}{a^3}\frac{\partial^{2}}{\partial\phi^{2}} -6\frac{c}{\kappa}\tilde{k}a + ca^{3}V(\phi)\right)\psi(a,\phi)=0
\label{WD}
\end{equation}
with the real numbers $i,j,k$ fulfilling $i+j+k=1$. This is our approach to the factor ordering problem. It is easy to show that the factor ordering problem is impossible to resolve by means of the semiclassical limit, because by assuming $\psi=\exp{\frac{\mathrm{i}}{\hbar}}S$ one obtains a Hamilton-Jacobi equation without regard to the numbers $i,j,k$ after neglecting terms containing $\hbar$.
\par
There are some orderings which one can prefer. Most popular is to consider expressions with derivatives as a Laplace-Beltrami (LB) operator \cite{dewitt} in superspace. There is only one such ordering: $i=2,j=-1$, $k=0$ ie. $\frac{1}{a^3}\hat{\pi}\hat{\pi}$. This is denoted as ``D'Alembert operator ordering" in \cite{KW}. Another choice, namely $i=k=0$, $j=1$, was used by Vilenkin \cite{V}.
\par
On the other hand, there exists an infinity of possibilities of ordering. We will show solutions for a massless field in all these cases. Nevertheless, someone might prefer special selections: for example all symmetric orderings of $\hat{a}$ and $\hat{\pi}$ are
\begin{equation}
\frac{\pi^2}{a^3} \quad\longmapsto\quad  -\frac{1}{a^{\frac{1}{2}(3-n)-1}} \frac{\partial}{\partial a} \frac{1}{a^{n-1}} \frac{\partial}{\partial a}  \frac{1}{a^{\frac{1}{2}(3-n)}},
\label{sus}
\end{equation}
where $n-1=j$,
or in terms of the canonical operators $\hat{a}$ and $\hat{p}_{a}$
\begin{equation}
\frac{p^2}{a} \quad\longmapsto\quad  -\frac{1}{a^{\frac{1}{2}(1-j)}} \frac{\partial}{\partial a} \frac{1}{a^{j}} \frac{\partial}{\partial a}  \frac{1}{a^{\frac{1}{2}(1-j)}}.
\label{su}	
\end{equation}
This operator is not selfadjoint on the Hilbert space mentioned before, but on $\mathcal{L}^2(\mathbb{R}^+,{\rm d}a;\mathbb{R},{\rm d}\phi)$.
A complex generalization (here i is the imaginary unit) of a special case of (\ref{sus}) is
\begin{equation}
\frac{\pi^2}{a^3} \quad\longmapsto\quad \frac{1}{2}\left(\frac{1}{a^3}\hat{\pi}\hat{\pi}+\hat{\pi}\hat{\pi}\frac{1}{a^3}\right)=\frac{1}{a^{\frac{3}{2}(1\pm\mathrm{i})}}\hat{\pi}a^{\pm3\mathrm{i}}\hat{\pi}\frac{1}{a^{\frac{3}{2}(1\pm\mathrm{i})}}.
\label{sus2}
\end{equation}
Also this last choice represents manifestly selfadjoint operators on $\mathcal{L}^{2}(\mathbb{R}^+,\frac{da}{a})$.
\par
For the massless scalar field the WD equation (\ref{WD}) is simply solvable for every ordering and curvature, because one obtains an equation with separable variables. An Ansatz $\psi(a,\phi)=A(a)\varphi(\phi)$ leads to
\begin{eqnarray}
\label{rovniceproA}
 \frac{1}{a^{i}} \frac{\rm d}{{\rm d} a} \frac{1}{a^{j}} \frac{\rm d}{{\rm d}a}  \frac{1}{a^{k}}A(a) - \frac{144}{(8\pi l^2_p)^2}\tilde{k}aA(a) + \frac{\lambda^{2}}{a^{3}}  A(a) &= & 0,\\
\label{rovniceprofi}
\frac{{\rm d}^{2}\varphi(\phi)}{{\rm d}\phi^{2}} + \frac{\lambda^{2}\kappa}{6}\varphi(\phi) &= & 0,
\end{eqnarray}
here $l_p$ is the Planck length. The solution of the second equation is a plane wave $\varphi\approx e^{\pm \mathrm{i} p_\phi \phi}$, which corresponds to the result of the classical theory, where the momentum of the scalar field is an integral of motion. 
\par
Now we introduce a new variable $b=\frac{3}{4\pi}\left( \frac{a}{l_p} \right)^2$ and a new function $B(b)$ as $A(a)=a^{1+\frac{k-i}{2}}B(b(a))$, so equation (\ref{rovniceproA}) takes the form
\begin{equation}
B^{\prime\prime}+\frac{B^{\prime}}{b}-\left(\tilde{k}+\frac{\nu^2}{b^2}\right)B=0
\label{bes}
\end{equation}
with 
\begin{equation} 
\nu^2 \equiv \frac{1}{16}\left[(j+1)^2-4\lambda^2\right].
\label{eps1}
\end{equation}
$\nu$ may be real or imaginary in dependence on the relation between the energy or momentum of the field $\phi$, given by $\lambda=\sqrt{\frac{6}{\kappa}}\,p_\phi$, and the ordering parameter $j$.
In all the cases we can write the $a$-dependent part of the wave function as
\begin{equation}
A(a)=\left(\scriptstyle \sqrt{\frac{3}{4\pi}} \frac{a}{l_p}\right)^{1+\frac{k-i}{2}} B_{\nu}\left(\scriptstyle \frac{3}{4\pi}\left(\frac{a}{l_p}\right)^2\right)
\label{ousvSI}
\end{equation}
with $B_\nu$ being a solution of (\ref{bes}).

\section{Inner product}
If we introduce the notation $x^\alpha=(x^1=a,x^2=\phi)$, we will get after some calculations from (\ref{WD})

\begin{equation}
\partial_ \alpha J^\alpha=\frac{\kappa}{24}\frac{i-k}{a^2}\left(\frac{\partial \psi^*}{\partial a}\psi -\psi^* \frac{\partial \psi}{\partial a}\right),
\label{div}
\end{equation}
with the flow vector
\begin{equation}
J_\alpha:= \left(\kappa\left(\psi\frac{\partial\psi^*}{\partial a}-\psi^*\frac{\partial\psi}{\partial a}\right),\;  \psi \frac{\partial \psi^*}{\partial \phi}-\psi^*\frac{\partial \psi}{\partial \phi} \right).
\label{flow}
\end{equation}
The metric in $x^\alpha$ space is $G_{\alpha\beta}={\rm diag}(-24a,4a^3)$, which follows directly from (\ref{HV}). If we choose $i=k$, then equation (\ref{div}) 
represents an equation of continuity for the flow $J_{\alpha}$. Of course, the right hand term of (\ref{div}) and the first term of (\ref{flow}) vanish, if the $a$-dependent part of the wave function $\psi$ is a real function of a real variable. This holds for the closed model, as we shall see in the next section. 

In the symmetric case $i=k$ and in the case of a real wave function of $a$ a  volume integral of (\ref{div}) in two-dimensional superspace spanned by $a$ and $\phi$ leads to an integral over the flow through some closed loop (= one-dimensional closed surface) in (two-dimensional) superspace, which is equal to zero. Moreover, in the closed case there is only one nonzero component of $J^\alpha$, namely $J^2$, flowing in the direction of $\phi$. Thus, if $\phi$ is considered as internal time, for the closed model the integral over $a$ is conserved in time and represents an indefinite norm corresponding to the inner product $\langle\psi,\varphi\rangle=\int\,\frac{{\rm d}a}{a}(\psi^*\partial_\phi \varphi -\varphi\,\partial_\phi \psi^*)$. (As long as we have a certain value of $p_\phi$, or the parameter $\lambda$, respectively, the role of the internal time $\phi$ is analogous to the role of time in an energy eigenstate in quantum mechanics -- there is no dynamical evolution. Dynamics arises when we take superpositions into account.)

The sign of $p_{\phi}$ is arbitrary because the conservation law holds for the positive as well as for the negative sign of $p_{\phi}$ (if $i=k$) and, as long as $p_\phi\neq0$ and as there are no superpositions of wave functions with positive and negative values of $p_\phi$, we can interprete $|\psi^*\partial_\phi\psi-\psi\partial_\phi\psi^*|$ as probability density, which will be seen to be in accordance with the classical one.
\par
The requirement of writing (\ref{WD}) in covariant form, i.\,e. with the derivatives in form of the LB operator $\frac{1}{a}\hat\pi\hat\pi$ mentioned above, leads to the formula
\begin{equation}
\nabla_\alpha J^\alpha=\frac{1}{\sqrt{-G}}\,\partial_\alpha (\sqrt{-G}J^\alpha)=0,
\label{div2}
\end{equation}
with the same $J_\alpha$ as in (\ref{flow}) and with an interpretation as probability density for $\sqrt{-G}J_\alpha$ rather than for the components $J_{\alpha}$.
\par
Now we can integrate the equation (\ref{div2}) over a volume in superspace and using the Gau\ss\ integral theorem, which holds in all dimensions, we get

\begin{equation}
\int \nabla_\alpha J^{\alpha}\sqrt{-G} \; \mathrm{d}^{2} x=\int\partial_{\alpha}(\sqrt{-G} J^{\alpha}) \mathrm{d}^{2}x =\int \sqrt{-G}J_{\alpha}n^{\alpha} \mathrm{d}l,
\end{equation}
where $\mathrm{d}l$ is the line element of the closed loop and $n^{\alpha}$ is the normal to this loop. The conserved quantity is $\sqrt{-G} J_{\alpha}$. The DeWitt condition $A(a\rightarrow 0)$ is postulated for a function which has an interpretation as a probability density. This condition means that the classical singularity at $a=0$ is avoided. In our case it means, that $|J_2(a\rightarrow0)|=0$ for the symmetrical ordering or $|\sqrt{-G}J_2(a\rightarrow0)|=0$ for the LB ordering.
But we also must study whether the right hand side integral is correctly defined. The overlap of these two conditions also restricts the factor ordering parameters.

\section{Closed model}
\subsection{Solutions of the hamiltonian constraint}
If we choose $\tilde{k}=1$, we get from (\ref{bes}) a modified Bessel equation of the order $\nu$, which can be viewed as a Bessel equation with purely imaginary argument. The general solution has the form
\begin{equation}
B(b)=c_{1} I_{\nu}(b)+c_{2}K_{\nu}(b).
\label{beprouz}
\end{equation}
Here $I_{\nu}(b) \equiv \mathrm{i}^{-\nu}J_{\nu}(\mathrm{i}b)$. This function is not limited in the asymptotic domain and for this reason it is not normalizable. The second function is a Basset function which is defined as

\begin{equation}
K_{\nu}(b)=\frac{\pi}{2}\mathrm{i}e^{\frac{\pi}{2} \nu \mathrm{i}} H_{\nu}^{1}(\mathrm{i}b).
\label{basset}
\end{equation}

Let's first assume that the order $\nu$ is imaginary and write $\nu={\rm i}\varepsilon$. This is the case when the square of $\lambda$ is above a critial value, coming from (\ref{eps1}),
\begin{equation} \label{krit}
\lambda^2>\lambda^2_{\rm cr}=\frac{(j+1)^2}{4}.
\end{equation}
In terms of the absolute value of the canonical momentum this means
\begin{equation}
|p_\phi|>p_{\rm cr}={\scriptstyle\sqrt{\frac{\hbar\kappa}{6}}\,\frac{1}{2}}\,|j+1|.
\end{equation}

If we put $b\equiv x^2$, we can rewrite the Basset function of imaginary order in the form

\begin{equation}
K_{\mathrm{i}\varepsilon}(x^2) = \frac{\pi}{\sinh \pi\varepsilon} \sum_{m=0}^{\infty} \left( \frac{x^2}{2} \right)^{2m} \left[\frac{\beta \cos\left(\varepsilon\ln \frac{x^2}{2}\right) -\alpha \sin\left(\varepsilon \ln\frac{ x^2}{2}\right)}{\alpha^2 + \beta^2}\right],
\label{basset2}
\end{equation}
where $\alpha=\alpha(m,\varepsilon)$ is the real and $\beta=\beta(m,\varepsilon)$ is the imaginary part of $\Gamma(1+m)\Gamma(1+m+\nu)$. 
From this it follows that the Basset 
function of pure imaginary order and with arguments from $\mathbb R^+$ is a real function of a real variable. In the asymptotic limit Basset functions of real or imaginary order reduce to the form 

\begin{equation}
K_{\nu}(b)\simeq\sqrt{\frac{\pi}{2b}}\:e^{-b}.
\end{equation}
On the other hand the function is limited for $x^2\rightarrow 0_+$, as can be seen from (\ref{basset2}). For these reasons we will accept the Basset function of imaginary order as the appropriate part of (\ref{beprouz}). So the physically acceptable solution of the Wheeler-DeWitt equation (\ref{rovniceproA}) in the closed case, for the momentum of the scalar field above the critical value, is 
\begin{equation}
A_{\lambda}(a)=c(\lambda) {\scriptstyle\left(\sqrt{\frac{3}{4\pi}} \frac{a}{l_p}\right)}^{1+\frac{k-i}{2}} K_{\mathrm{i}\varepsilon{\scriptscriptstyle(\lambda)}} {\scriptstyle \left(\frac{3}{4\pi}\left(\frac{a}{l_p}\right)^2\right)}.
\label{reseniprouz}
\end{equation}

\subsection{Inner product}
Taking the scalar product in the Hilbert space $\mathcal{L}^2(\mathbb{R}^+,\frac{{\rm d}a}{a};\mathbb{R},{\rm d}\phi)$ from the foregoing section, we arrive, up to a $\lambda$-dependent factor, at the following norm of a mode $A_\lambda(a)$
\begin{equation}
\int_0^\infty\frac{{\rm d}a}{a}\,a^{2+k-i}\,K_{{\rm i}\varepsilon}^2\scriptstyle\left(\frac{3}{4\pi}\left(\frac{a}{l_p}\right)^2\right).
\end{equation}
This is finite if $2+k-i>0$. If $2+k-i>1$, then the integrand goes to zero for $a\rightarrow0$. Interpreted as probability density to find the model at a certain value of $a$, the DeWitt condition of the avoidance of the classical singularity $a=0$, is satisfied. The above conditions depend on the factor ordering parameters $i$ and $k$; for the selfadjoint choice (\ref{sus}), where $k-i=1$, they are satisfied.

The alternative choice (\ref{su}) with $i=k$ and the corresponding integration measure d$a$ in the Hilbert space leads to exactly the same result. For both choices normalizability restricts factor ordering, the DeWitt condition enhances the restriction. Figure 2 compares the assumed quantum probability density with the classical one, obtained from (\ref{apunkt}) as the inverse of the velocity $\dot a$,
\begin{equation}
p_{\rm cl}=\frac{1}{\dot a}=\frac{a^2}{\sqrt{a_m^4-a^4}}.
\end{equation}  

In the case of undercritical field momentum, when $\lambda^2<\left(\frac{j+1}{2}\right)^2$, we obtain a Basset function of real order, $K_\nu\left(\frac{3}{4\pi}\left(\frac{a}{l_p}\right)\right)$, with
\begin{equation}
\nu=\frac{1}{4}\sqrt{(j+1)^2-4\lambda^2}
\end{equation}
as $a$-dependent part of the wave function. Obviously the domain of $\lambda$ of these solutions depends on the factor ordering parameter $j$, if $j=-1$, there is only one solution $K_0$ for $\lambda=0$. Given the case that $j\neq-1$ there are solutions for $\lambda \in (-\frac{j+1}{2},\frac{j+1}{2})$, but they may be normalizable or not. From the behaviour of the Basset functions of order $\nu\neq0$ for $a$ close to zero,
\begin{equation}
K_\nu\left(\mbox{$\frac{3}{4\pi}\left(\frac{a}{l_p}\right)^2$}\right)\sim a^{-2\nu},
\end{equation}
it follows that the normalizing integral $\int_0^\infty\frac{{\rm d}a}{a}\,A_\lambda^2(a)$
exists if
\begin{equation}\label{norm}
1+k-i-\sqrt{(j+1)^2-4\lambda^2}>-1.
\end{equation}
For the LB operator, for example, with $i=2$, $j=-1$, and $k=0$, the only solution of this type with $\nu=0$ is not normalizable. Here the normalizable solutions with imaginary $\nu$ for $p_\phi>0$ and those with $p_\phi<0$ are separated by one non-normalizable solution with $p_\phi=0$.  

A normalizability condition resulting from (\ref{norm}) is
\begin{equation}
i-k<2.
\end{equation}
If we consider factor ordering parameters close to this boundary, characterized by a small number $\epsilon$, $i-k=2-\epsilon$, we can express $i$ and $k$ by $j$ and a further parameter $\delta$, using $i+j+k=1$,
\begin{equation}
i=\frac{1}{2}(3-j-\delta), \hspace{2cm}k=\frac{1}{2}(-1-j+\delta).
\end{equation}
Now we insert back into (\ref{norm}) and obtain a lower boundary for $\lambda$,
\begin{equation}\label{epsilon}
\left(\frac{j+1}{2}\right)^2-\left(\frac{\delta}{2}\right)^2<\lambda^2.
\end{equation}

So, for overcritical values of $\lambda^2$, there are normalizable, oscillating wave functions, for undercritical values there may be, depending on the parameter $\delta=2-i+k$, normalizable non-oscillating functions for
\begin{equation} \lambda\in\left(-\frac{j+1}{2},-\frac{j+1}{2}+\frac{\delta}{2}\right)\cup\left(\frac{j+1}{2}-\frac{\delta}{2},\frac{j+1}{2}\right),
\label{n}
\end{equation}
Whether or not there are normalizable wave functions for undercritical values of $\lambda$, depends on the relation between the deviance of $i-k$ from 2 and $j$. The same applies for the ocurrence of non-normalizable functions, which seperate normalizable functions with positive and negative values of $\lambda$ from each other. Of course, for the symmetrical factor orderings the functions are normalizable.
 
 
\subsection{The DeWitt condition}
In terms of the field momentum we can separate solutions, which fulfil some continuity equation, into two parts. The first kind of states, for which holds

\begin{equation}
p_{\phi}\in \mathbb{R}\backslash\ {\scriptstyle \left(-\sqrt{\frac{\hbar^2 \kappa}{6}}\frac{1}{2}(j+1),\sqrt{\frac{\hbar^2 \kappa}{6}}\frac{1}{2}(j+1)\right)},
\label{omezeniimpulzupole}  
\end{equation}
will be called dynamical (overcritical). The remaining states will be called nondynamical (undercritical). We called these states dynamical, because close to the singularity the probability densities derived from these states are comparable with the classical ones. Moreover, dynamical states must be represented by a function oscillating near the classical singularity.  Effectively the classical quantity $1/\dot{a}\approx a^2 $ is comparable with the corresponding quantum expression near the classical singularity (for solutions fulfilling some continuity equation), as one can see from (\ref{basset2}) and (\ref{reseniprouz}). 
Unfortunately, the mathematically nice Laplace-Beltrami factor ordering has no nondynamical states for non zero field momentum (see (\ref{omezeniimpulzupole}) with $j=-1$). Any other factor ordering has nondynamical states for $p_{\phi}\neq 0$.
\par
 Every dynamical state, which fulfils a continuity equation, fulfils the DeWitt condition. Nondynamical states fulfil the DeWitt condition iff 

$$
p_{\phi}\in {\scriptstyle \left[-\frac{1}{2}\sqrt{\frac{\hbar^2 \kappa}{6}}(j+1),-\frac{1}{2}\sqrt{\frac{\hbar^2 \kappa}{6}}\sqrt{(j+1)^2-4}\right)\cup \left(\frac{1}{2}\sqrt{\frac{\hbar^2 \kappa}{6}}\sqrt{(j+1)^2-4}, \frac{1}{2}\sqrt{\frac{\hbar^2 \kappa}{6}}(j+1)\right]}.
$$

If $j=1$ the unification of dynamical and nondynamical states satisfies the DeWitt condition for momenta $p_{\phi}\in \mathbb{R} \backslash \left\{0\right\}$.
Note that the symmetric ordering $j=1$ is derived as semiclassical limit in loop quantum cosmology \cite{MB}. It has a maximal set of values of the field momentum, for which $J_2(a\rightarrow 0)=0$ holds. 
Any other choice fulfilling a continuity equation (i.e. symmetrical) doesn't fulfil the DeWitt condition in some finite momentum interval.

\begin{figure}[h]
\centering
\rotatebox{-90}{\scalebox{0.50}{\includegraphics{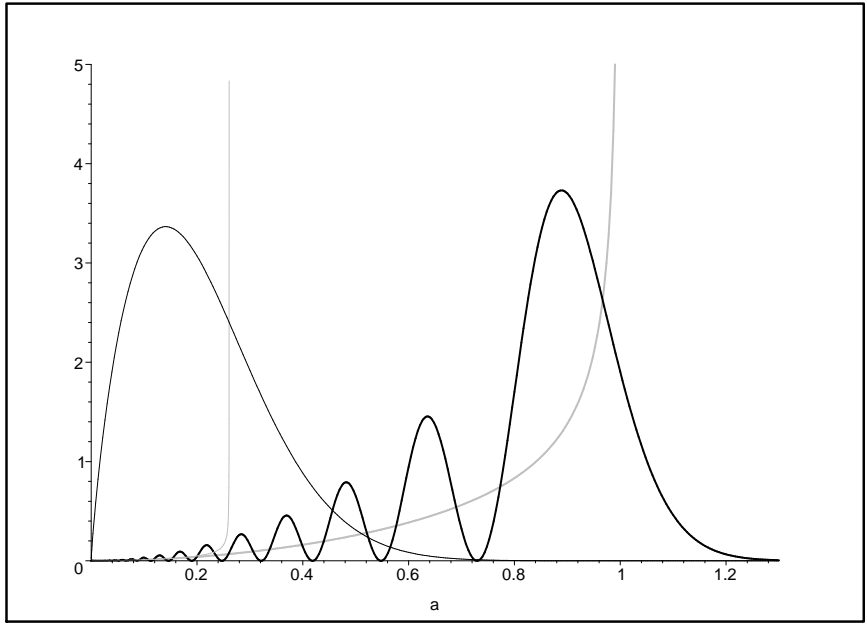}}}
\caption{Classical and quantum probability density.}
\label{closed}
\end{figure}

On figure \ref{closed} classical (grey) and quantum (black) probability densities for $j=1$ are displayed. We can see a density constructed for a dynamical (bold lines; $p_{\phi}=2\sqrt{6}$) and a nondynamical state (thin lines; $p_{\phi}=1/3$). 
\par
On the classical level there is no reason for a restriction of $\lambda$. This could be an argument in favour of a factor ordering with $j=-1$, $n=0$ or $j=\mp 3\mathrm{i}-1$. Specifically this holds for the LB operator $\frac{1}{a^3}\hat{\pi}\hat{\pi}$. On the other hand, in loop quantum cosmology  \cite{MB} other factor orderings are preferred and an analogous critical field energy (of the order of the Planck energy) for a dynamical evolution of the model exists. 
\par
Nevertheless, an argument in favour of the symmetrical (\ref{su}) or LB ordering is that for it we can correctly formulate the continuity equation. LB ordering may be interpreted as ordering of the selfadjoint operators $\frac{1}{a^3}\hat{\pi}\hat{\pi}$.   

\section{Solution for the open model}

In spite of principal problems with the canonical quantization of open cosmological models \cite{MC}, here we simply take over the Wheeler-DeWitt equation from the closed model and change the sign of the curvature and justify this step by the correct classical limit of the resulting wave function. 
For $\tilde k =-1$ (\ref{bes}) becomes a Bessel equation and as in the foregoing case we must distinguish between the cases $\nu^2>0$ and $\nu^2<0$. For $\nu={\rm i}\varepsilon$ in the case $\lambda^2>\lambda^2_{\rm cr}$ we get a solution in terms of two linearly independent Bessel functions of imaginary order,
\begin{equation}
B(b)=c_{1}J_{{\rm i}\varepsilon}(b) + c_{2}J_{-{\rm i}\varepsilon}(b),
\label{beprohyp}
\end{equation}
as before. (The two Bessel functions are complex conjugate. Alternatively the solution can be given in terms of the Hankel functions $H^{(1)}_{i\varepsilon}$ and $H^{(2)}_{i\varepsilon}$.)
The wave function for the open model becomes
\begin{equation}
A_{\lambda}(a)= {\scriptstyle\left(\sqrt{\frac{3}{4\pi}} \frac{a}{l_p}\right)}^{\scriptscriptstyle1+(k-i)/2} \left[c_1(\lambda)\,J_{\mathrm{i}\varepsilon{\scriptscriptstyle(\lambda)}} {\scriptstyle \left(\frac{3}{4\pi}\left(\frac{a}{l_p}\right)^2\right)}
+c_2(\lambda)\,J_{-{\rm i}\varepsilon{\scriptscriptstyle(\lambda)}}{\scriptstyle\left(\frac{3}{4\pi}\left(\frac{a}{l_p}\right)^2\right)}\right].
\label{reseniproot}
\end{equation}



For convenient assumptions about the factor ordering, i.e. $k-i=0$ for the Hilbert space with the measure d$a$, and $k-i=1$ for the measure d$a/a$, the square of the absolute value of the above functions approximates very closely the classical probability density $1/\dot a=a^2/\sqrt{a^4+p^4}$ from (\ref{minus1}), see fig. 4. 

\begin{figure}[ht]
\centering
\includegraphics{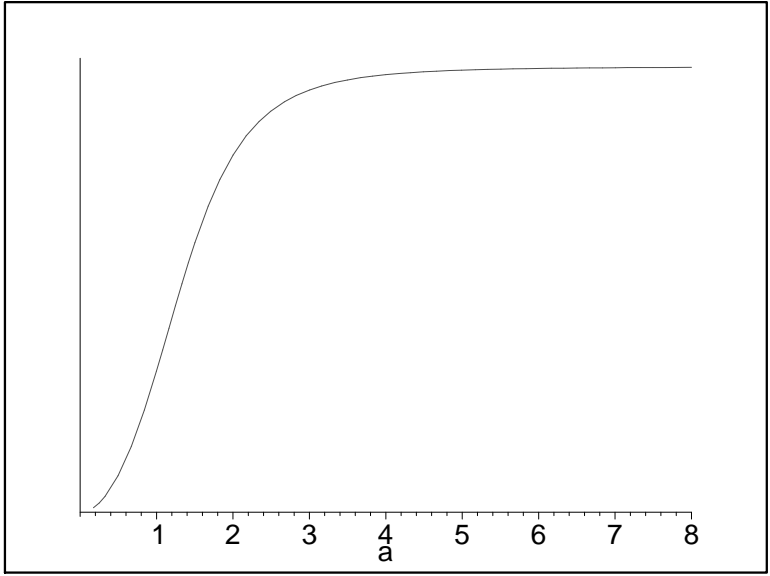}
\caption{\footnotesize Plot of $a^2\,|J_{{\rm i}\varepsilon}(a^2)|^2$. }
\end{figure}

For real $\nu$ the solution of (\ref{bes}) may be written as superposition of Hankel functions $H_\nu^{(1)}(b)$ and $H_\nu^{(2)}(b)$ or the corresponding Bessel functions of the first and second kind. In the first case only the expression $b|H^{(i)}_\nu(b)|^2$ as probability density would be constant for large $a$ and so display the classical behaviour, but it diverges at $a=0$ for $\nu\geq1/2$. If we admit only $J_\nu(b)$ for $\nu>0$ we have a  function that vanishes at the classical singularity, but it does not have the character of an outgoing or ingoing wave function for large $a$. Moreover, the absolute squares of the Bessel functions of real order are oscillating and do not approximate the classical behaviour shown in fig. 4.

In the spatially flat case, when $\tilde{k}=0$, the situation is analogous. For $\nu={\rm i}\varepsilon$ the solutions of (\ref{bes}) are $\sim a^{\pm2{\rm i}\varepsilon}$ and have a dynamical interpretation as in- and outgoing wavefunctions. Given an appropriate factor ordering with respect to the chosen Hilbert space measure, they are normalizable to a $\delta$-function. For real $\nu$ the solutions do not have these properties. 

\section{Summary}

In the spatially flat case there is a simple distinction betwee over- und undercritical values of the field momentum $p_\phi$, or the parameter $\lambda$, respectively: For too small values there is no dynamical wave function, corresponding to the classical solution. In the other cases there may be a nonvanishing range of the above parameters between the critical value and zero, in which the wave functiones might be physically acceptable. For $k=-1$ this means there are two linearly independent solutions $H^{(1)}$ and $H^{(2)}$ and a probability density, according to the classical limit. In the closed case $k=1$ we mean by acceptable wave functions normalizable ones. The existence and the extension of a range of undercritical energies with normalizable wave functions depends crucially on the factor ordering and is more restricted by classical considerations in the open case, where the limit of the probability density for large $a$ singles out symmetric orderings. A physical interpretation of ``undercritical" normalizable wave functions is still missing. 

The flow $J_{\alpha}$ has two nonvanishing components, but we reserve an interpretation as probability density for $|\sqrt{G}J_{2}|$ in the case of LB ordering, or for the component $|J_2|$ in the case of symmetric ordering. 

To compare once more with \cite{KW}, not the form of the potential, but the kinetic energy of the scalar field plays a role in our paper. Due to this, rather than ``tunneling" and ``oscillating" regions in superspace, i.e. a different behaviour of the wave function in different domains of the scale parameter, we obtain different regions concerning the energy of the field modes. Our result is not a rigid restriction of factor orderings or the picking-out of some distinguished ones, but rather a subtle dependence of the critical values of the field energy on the factor ordering. \\

\noindent {\bf Acknowledgement:} Supported by the Ministry of Education of the Czech Republic under the project MSM 0021622409

\end{document}